\documentclass[oldversion]{aa}
\usepackage{amsmath,graphicx,amssymb}
\usepackage{txfonts}
\usepackage{natbib}
\bibpunct{(}{)}{,}{a}{}{,}

\newcommand{\zav}[1]{\left(#1\right)}
\newcommand{\hzav}[1]{\left[#1\right]}

\newlength\staretab
\newcommand{\Teff}{\mbox{$T_\mathrm{eff}$}}

\makeatletter
\def\sgn{\mathop{\operator@font sgn}\nolimits}
\makeatother

\begin{document}

\title{Effect of rotational mixing and metallicity on the hot star wind mass-loss rates}

\author{J.~Krti\v{c}ka\inst{1} \and J. Kub\'at\inst{2}}

\institute{\'Ustav teoretick\'e fyziky a astrofyziky, Masarykova univerzita,
           Kotl\'a\v rsk\' a 2, CZ-611\,37 Brno, Czech
           Republic
           \and
           Astronomick\'y \'ustav, Akademie v\v{e}d \v{C}esk\'e
           republiky, CZ-251 65 Ond\v{r}ejov, Czech Republic}

\date{Received}

\abstract{Hot star wind mass-loss rates depend on the abundance of individual
elements. This dependence is usually accounted for assuming scaled solar
chemical composition. However, this approach may not be justified in evolved
rotating stars. The rotational mixing brings CNO-processed material to the
stellar surface, increasing the abundance of nitrogen at the expense of carbon
and oxygen, which potentially influences the mass-loss rates. We study the
influence of the modified chemical composition resulting from the rotational
mixing on the wind parameters, particularly the wind mass-loss rates. We use our
NLTE wind code to predict the wind structure and compare the calculated wind
mass-loss rate for the case of scaled solar chemical composition and the
composition affected by the CNO cycle. We show that for a higher mass-fraction
of heavier elements $Z/Z_\odot\gtrsim0.1$ the change of chemical composition
from the scaled solar to the CNO-processed scaled solar composition does not
significantly affect the wind mass-loss rates. The missing line force caused by
carbon and oxygen is compensated for by nitrogen line force. However, for a very
low-mass fraction of heavier elements $Z/Z_\odot\lesssim0.1$ the rotational
mixing significantly affects the wind mass-loss rates. Moreover, the decrease of
the mass-loss rate with metallicity is stronger at such low metallicities. We
study the relevance of the wind momentum-luminosity relationship for different
metallicities and show that for a metallicity $Z/Z_\odot\lesssim0.1$ the
relationship displays a large scatter, which depreciates the use of this
relationship at the lowest metallicities.}

\keywords {stars: winds, outflows -- stars:   mass-loss  -- stars:
early-type -- hydrodynamics}

\titlerunning{The effect of rotational mixing and metallicity on hot star wind mass-loss
rates}

\authorrunning{J.~Krti\v{c}ka and J. Kub\'at}
\maketitle

\section{Introduction}

With the advent of new 8m class telescopes and space-borne instruments the
attention of stellar astronomers extends from our Galaxy to galaxies in the
Local Group. Such observations do not only provide population studies of stars
at a comparable distance, but enable us to study the influence of another key
evolutionary parameter in addition to the stellar mass, namely the different
chemical composition of stars, which is usually referred to as the stellar
metallicity. In the hot star domain it was possible to study stars from
metallicities similar to the Galactic ones in M33 \citep{lucka} to stars with
metallicities of a fraction of a Galactic one found in the dwarf irregular
galaxies of the Local Group \citep{tramp,hergagalic}.

The influence of metallicity is especially important for hot star winds, which
are the key mass-loss rate mechanism in hot stars. Stellar winds of hot stars
are accelerated mainly as a result of light absorption in numerous lines of
heavier elements including carbon, nitrogen, oxygen, and iron \citep{abb82}.
Consequently, in addition to the stellar luminosity, the metallicity is the most
important parameter determining the amount of stellar mass lost per unit of
time, that is, the mass-loss rate $\dot M$. For example, the absence of elements
heavier than helium (hereafter metals) completely inhibits any line-driven
outflow in massive first stars \citep{bezvi}, while the mass-loss rate in
solar-metallicity massive stars amounts typically to up to
$10^{-6}\,M_\odot\,\text{year}^{-1}$ \citep[e.g.,][] {pulvina} or even more.

The effect of metallicity on the hot star wind was studied in detail for a wide
range of metallicities \citep{vikolamet,lijana,pakul}, providing mass-loss rate
as a function of metallicity $Z$. Due to the lack of suitable abundance model,
most of theoretical work used the solar chemical composition with abundances
multiplied by the same factor for all metals (hereafter the scaled solar
chemical composition), yielding the dependence of mass-loss rate on metallicity
as $\dot M\sim Z^{0.67}$ \citep{nlteii}. However, there is no reason to believe
that the scaled solar chemical composition is always appropriate. It may
lead to wrong results, especially at low metallicities. Complex chemical
evolution of galaxies can lead not only to deviations from scaled solar chemical
composition, but the rotational mixing in stars may change the surface chemical
composition even during their main-sequence evolution \citep{zen002}.

In nonrotating main-sequence stars the diffusion processes are rather weak, and
the diffusion time is too long to bring the fresh elements that are synthesized
in the stellar core to the stellar surface in large amounts \citep[for clues to
additional mixing see, e.g.,][]{boulamama}. However, in rotating stars the
effects of instabilities caused by nonuniform rotation and meridional
circulation may imprint the signatures of thermonuclear reactions on the surface
chemical composition \citep{memarot}. In rotating first stars, the occurrence of
metals on the stellar surface as a result of mixing may trigger the line-driven
outflow \citep{cnovit}. At higher initial metallicity the change of fraction of
CNO elements is the most notable abundance effect of thermonuclear reactions.
The typical time of individual reactions of CNO cycle is different, the reaction
involving $^{14}$N being the slowest \citep{biblerot}. This typically leads to
the transition from the abundance with oxygen as the third most abundant element
to nitrogen-rich environment. Signs of such processes were detected from
spectroscopic analysis \citep{lovec,prfin}. 

Changes in abundances affect the radiative force. The CNO elements do not
contribute to the radiative force by the same amount, and their contribution
also depends on the temperature \citep[e.g.,][]{vikolamet,cnovit}. Consequently,
two stars in the same evolutionary phase with the same initial chemical
composition may have different wind mass-loss rates purely as a result of
rotational mixing. These processes were not studied theoretically in detail.
Here we provide a study of the effect of chemical composition modified by the
mixing on the hot star wind mass-loss rates.

An ideal way how to account for this effect would be to calculate a wind model
at each time-step of the evolutionary models with proper surface abundances
\citep[as suggested by][]{lijana}. However, this would be prohibitively time
consuming. Therefore, to understand the effects of mixing on the mass-loss
rates, we provide wind models with scaled solar chemical composition in
comparison with models calculated assuming the same chemical composition, but
with CNO abundances modified as a result of the CNO cycle.

\section{Wind models}

We used the recent version of our NLTE wind models \citep{cmf1} with a radiative
force calculated consistently from the solution of the comoving frame (CMF)
radiative transfer equation with actual opacities. Our models are able to
predict wind mass-loss rates from the stellar parameters, that is, the effective
temperature, mass, radius, and \emph{arbitrary} chemical composition. Our
mass-loss rate is just a result of model calculations, we do not use any free
parameters to fix it. For the model calculation we assumed stationary and
spherically symmetric wind flow. 

The ionization and excitation state of the considered elements was calculated
from the equations of statistical equilibrium (NLTE equations). Models of
individual ions are based on the Opacity and Iron Project data
\citep{topt,zel0}. Part of the ionic models was adopted from the TLUSTY model
stellar atmosphere input data \citep{ostar2003,bstar2006}. For phosphorus we
employed data described by \citet{pahole}. The emergent surface flux was taken
from the H-He spherically symmetric NLTE model stellar atmospheres of \citet[and
references therein]{kub}.

The radiative force was evaluated using radiative flux calculated with the
solution method of the CMF radiative transfer equation following \citet{mikuh}.
The corresponding line opacity data were extracted in 2002 from the VALD
database (Piskunov et al. \citeyear{vald1}, Kupka et al. \citeyear{vald2}). To
calculate the radiative force and radiative cooling and heating terms
\citep{kpp} we used occupation numbers derived from the statistical equilibrium
equations. The hydrodynamical equations, that is, the continuity equation,
equation of motion, and the energy equation, were solved iteratively to obtain
the wind density, velocity, and temperature structure. The derived mass-loss
rate corresponds to the only smooth transsonic solution.

\begin{table}
\caption{Stellar parameters of the model grid}
\centering
\label{ohvezpar}
\begin{tabular}{rccrc}
\hline
\hline
& Model &$\Teff$ & $R_{*}$ & $M$  \\
& & $[\text{K}]$ & $[{R}_{\odot}]$ & $[{M}_{\odot}]$ \\
\hline main
& 300-5 & 30\,000 & 6.6 & 12.9 \\ sequence (V)
& 325-5 & 32\,500 & 7.4 & 16.4 \\
& 350-5 & 35\,000 & 8.3 & 20.9 \\
& 375-5 & 37\,500 & 9.4 & 26.8 \\
& 400-5 & 40\,000 &10.7 & 34.6 \\
& 425-5 & 42\,500 &12.2 & 45.0 \\
\hline giants
(III)
& 300-3 & 30\,000 &13.1 & 19.3 \\
& 325-3 & 32\,500 &13.4 & 22.8 \\
& 350-3 & 35\,000 &13.9 & 27.2 \\
& 375-3 & 37\,500 &14.4 & 32.5 \\
& 400-3 & 40\,000 &15.0 & 39.2 \\
& 425-3 & 42\,500 &15.6 & 47.4 \\
\hline supergiants (I)
& 300-1 & 30\,000 &22.4 & 28.8 \\
& 325-1 & 32\,500 &21.4 & 34.0 \\
& 350-1 & 35\,000 &20.5 & 40.4 \\
& 375-1 & 37\,500 &19.8 & 48.3 \\
& 400-1 & 40\,000 &19.1 & 58.1 \\
& 425-1 & 42\,500 &18.5 & 70.3 \\
\hline
\end{tabular}
\end{table}

The grid of stellar parameters corresponds to O stars in the effective
temperature range $30\,000-42\,5 00\,\text{K}$. The stellar masses and radii in
Table~\ref{ohvezpar} were calculated using relations of \citet{okali}. The solar
composition abundance ratios $\zav{N_\text{el}/N_\text{H}}_\odot$ for each
element ($\text{el}$) were taken from \citet{asp09}\footnote{For CNO elements
$\zav{N_\text{C}/N_\text{H}}_\odot=2.69\times10^{-4}$,
$\zav{N_\text{N}/N_\text{H}}_\odot=6.76\times10^{-5}$, and
$\zav{N_\text{N}/N_\text{H}}_\odot=4.9\times10^{-4}$.}. For the metal-deficient
composition we used the scaled solar mixture $N_\text{el}/N_\text{H}=Z/Z_\odot
\zav{N_\text{el}/N_\text{H}}_\odot$. The effect of the CNO cycle on the surface
abundance was taken into account by modifying the scaled solar mixture in such a
way that the total number density of CNO elements remained the same and the
fraction of carbon and oxygen relative to nitrogen are
$N_\text{C}/N_\text{N}=0.033$ and $N_\text{O}/N_\text{N}=0.1$ \citep[Table 25.3
therein]{biblerot}. Hereafter this chemical composition is referred to as the
CNO-processed scaled solar composition (see Table \ref{cnoab} for the adopted
abundances of CNO elements).

\section{Influence of modified abundance caused by the
CNO cycle on the mass-loss rates}

\begin{figure}
\centering
\resizebox{\hsize}{!}{\includegraphics{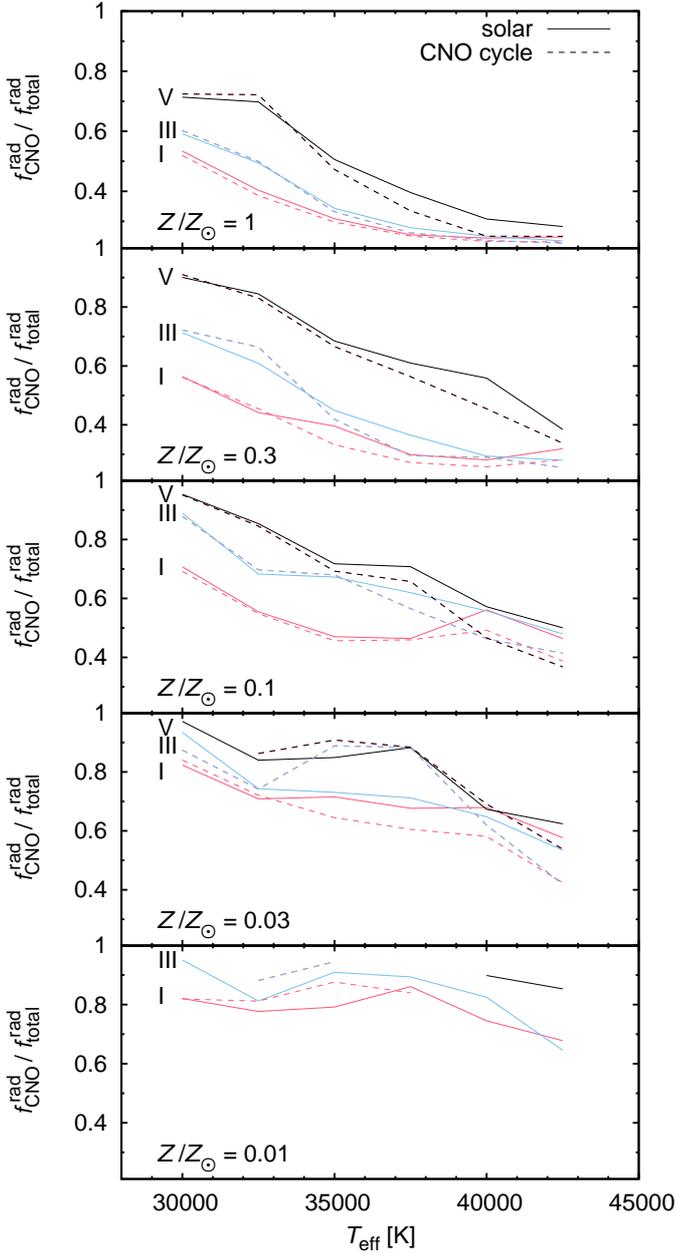}}
\caption{Fraction
of the line force caused by CNO elements (at the critical
point) as a function of the stellar effective temperature. Individual graphs are
plotted for different mass fractions of heavier elements $Z/Z_\odot$ given in
each graph. Each curve is labeled with the corresponding luminosity class.
Graphs are plotted for the scaled solar chemical composition (solid
lines) and for CNO-processed scaled solar chemical composition
(dashed lines).}
\label{cnosil}
\end{figure}

\begin{figure}
\centering
\resizebox{\hsize}{!}{\includegraphics{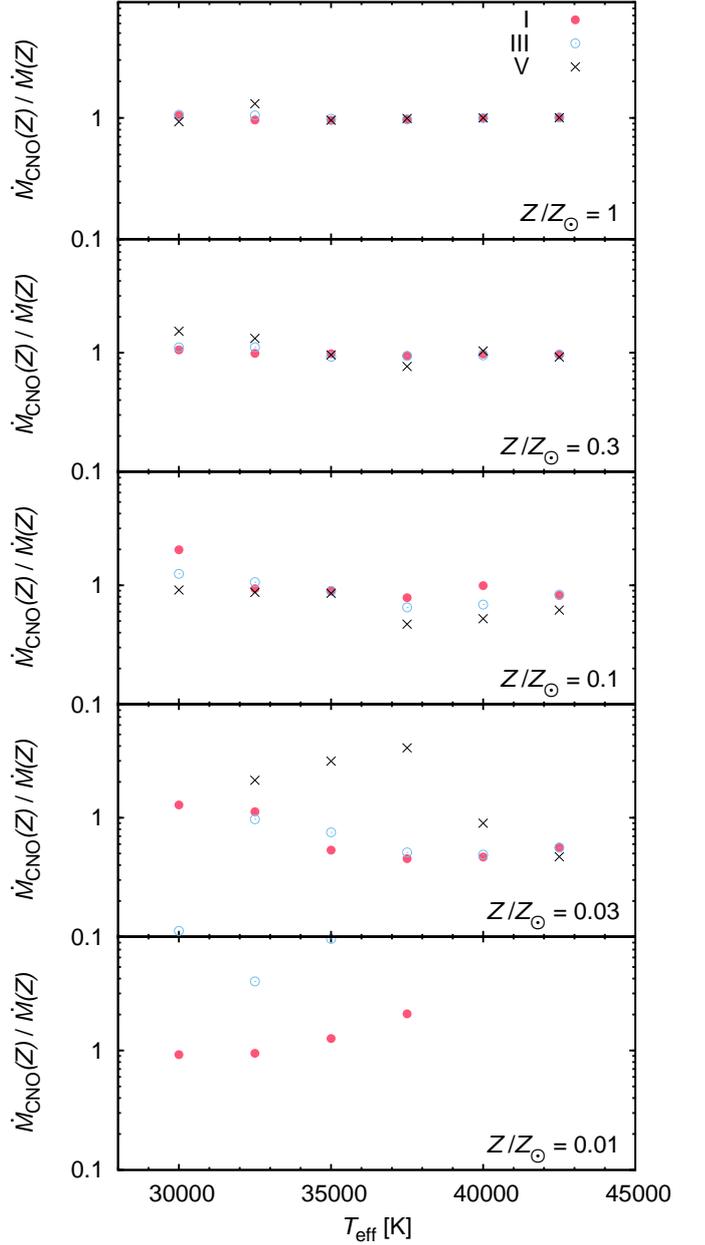}}
\caption{Ratio of the mass-loss rates calculated for a CNO-processed
scaled solar chemical composition
and for a scaled solar chemical composition as a
function of the effective temperature. Different symbols correspond to
different luminosity types as indicated in the uppermost panel.}
\label{ven}
\end{figure}

\subsection{Contribution of CNO elements to the radiative force}

In Fig.~\ref{cnosil} we plot the relative contribution of the CNO elements to
the total line radiative force. These graphs (and the detailed discussion) are
concerned with the radiative force at the wind critical point. The mass-loss
rate is determined by conditions below this point. Basic trends in
Fig.~\ref{cnosil} can be understood by considering that the radiative force
originating in optically thick lines is independent of the abundance of a given
element \citep{pusle,vikolamet}. Many lines of CNO and heavier elements are
optically thick in winds with high mass-loss rates (high densities). Heavier
elements (especially iron) have effectively more optically thick lines than CNO
elements, consequently, they dominate the radiative force in winds with high
density. At lower wind densities (i.e. for cooler stars with higher surface
gravities), the lines of less abundant elements (for example, of iron) become
optically thin and do not significantly the contribute to the radiative force.
On the other hand, more abundant CNO elements dominate line driving at low wind
densities, because their strongest lines remain optically thick even in such
conditions. Consequently, for a given metallicity the relative contribution of
CNO elements to the radiative force decreases with increasing stellar luminosity
and increasing effective temperature (increasing mass-loss rate). With
decreasing metallicity the radiative force decreases, and therefore the
mass-loss rate decreases. This leads to the increase of the relative
contribution of CNO elements to the radiative force. Hence CNO lines become very
important for the properties of the stellar wind in low-metallicity stars.

\subsection{Influence of abundance changes of CNO on the mass-loss rates}

In Fig.~\ref{ven} we provide the resulting ratio of the mass-loss rate for
CNO-processed scaled solar chemical composition compared with the mass-loss rate
predicted for scaled solar chemical composition (see also
Table~\ref{venzettab}). This ratio is a non-trivial function of the metallicity,
stellar effective temperature, and luminosity type. The complicated dependence
can be understood to be the result of the varying contribution of individual CNO
elements to the radiative force. The metallicity is the key parameter that
determines this dependence (see Table~\ref{esence}).

\begin{table*}
\caption{Change of the wind mass-loss rate $\dot M$ for CNO-processed chemical
composition}
\centering
\label{esence}
\begin{tabular}{cp{162mm}} 
\hline
$Z/Z_\odot$ & Effect\\
\hline
1 & contribution of CNO to the radiative force 20 -- 70\,\%, no
   significant change of $\dot M$ for CNO-processed composition\\
0.3& contribution of CNO to the radiative force 30--90\,\%, slightly
  higher $\dot M$ for models 300-5 and 325-5, slightly lower $\dot M$ for
  model 375-5, otherwise no effect\\
0.1& contribution of CNO to the radiative force $>40$\,\%, higher
  $\dot M$ for models 300-1 and 300-3, lower $\dot M$ for
  $T_\text{eff}\geq37\,500\,\text{K}$ (except 400-1)\\
0.03& contribution of CNO to the radiative force $>40$\,\%, higher
  $\dot M$ for models 325-5, 350-5, and 375-5, lower $\dot M$ for giant
  and supergiant models with $T_\text{eff}\geq37\,500\,\text{K}$ and for
  models 300-3 and 425-5\\
0.01& contribution of CNO to the radiative force $>60$\,\%, higher
  $\dot M$ for models 325-3, 350-3, and 375-1, no wind for
  $T_\text{eff}\geq 40\,000\,\text{K}$ for CNO-processed composition\\
\hline
\end{tabular}
\end{table*}

\paragraph{${Z=Z_\odot}$}

In most cases, the contribution of CNO elements to the radiative force is
relatively low for $Z=Z_\odot$. Moreover, many CNO lines have such high optical
depths that they are optically thick in the wind for the case of solar and
CNO-processed chemical compositions, and instead of weaker optically thick CO
lines for solar chemical composition, other N optically thick lines appear for
the CNO-processed chemical composition. The latter effect is important
especially for cooler stars from our sample. Consequently, the wind mass-loss
rates for solar chemical composition and for CNO-processed chemical composition
are basically the same.

\paragraph{${Z=0.3\,Z_\odot}$}

For giant and supergiant wind models at $Z=0.3\,Z_\odot$ the deficiency of the
radiative force caused by CO elements is compensated for by the radiative force
caused by N lines for CNO-processed chemical composition. The driving by iron
lines is significant as well. Consequently, the wind mass-loss rates for scaled
solar chemical composition and for CNO-processed chemical composition are almost
the same. For main-sequence wind models at lower effective temperatures
($T_\text{eff} \le 35\,000\,\text{K}$, models 300-5, 325-5, and 350-5) the
change to the CNO-processed chemical composition leads to a significant increase
of the radiative force caused by N lines, whereas in the hotter model 375-5 this
increase is not able to compensate for the decrease of the radiative force
caused by the decreasing contribution by CO elements. Consequently, the
main-sequence wind mass-loss rates are higher in models 300-5 and 325-5 with
CNO-processed chemical composition than for the solar mixture, whereas the
opposite is true for model 375-5. For $T_\text{eff}\geq40\,000\,\text{K}$ the
wind enters the high mass-loss rate (high-density) regime, therefore the change
of the fraction of individual CNO elements does not significantly affect the
radiative force and the mass-loss rate.

\paragraph{${Z=0.1\,Z_\odot}$}

For model 300-1 at $Z=0.1\,Z_\odot$ the mass-loss rate with CNO-processed
chemical composition is higher by about a factor of 2 than for the scaled solar
chemical composition. This is caused by a shift in ionization balance, which
increases the fraction of lower ions of oxygen and also subsequently of silicon
and sulfur. For $T_\text{eff}\geq37\,500\,\text{K}$ at $Z=0.01$ many strong
oxygen lines become optically thin for the CNO-processed chemical composition,
consequently, the mass-loss rate for the scaled solar chemical composition is
higher by about a factor of 2 than for the CNO-processed chemical composition.
The exception is model 400-1, where  the \ion{O}{iv} lines significantly
contribute to the line driving for the CNO-processed chemical composition.

\paragraph{${Z=0.03\,Z_\odot}$}

For giants and supergiants at $Z=0.03\,Z_\odot$ the oxygen lines that were
optically thick for the scaled solar chemical composition become optically thin
for the CNO-processed chemical composition and the nitrogen lines are not able
to compensate for the missing line force. For all models with
$T_\text{eff}=32\,500\,\text{K}$, the ionization fraction of \ion{N}{iv} becomes
high and the line force caused by strong oxygen lines for scaled solar chemical
composition is compensated for by \ion{the N}{iv} line force. In model 300-3 all
driving lines cease to be optically thick in the region of the critical point
for the CNO-processed chemical composition, consequently, the mass-loss rate
becomes very low. In the main-sequence model 375-5 at this metallicity the
\ion{O}{v} resonance line is optically thick. For the CNO-processed chemical
composition the \ion{N}{v} resonance doublet also becomes optically thick,
leading to a significant increase of the radiative force. In models 325-5 and
350-5 this effect is weaker, because \ion{C}{iv} and \ion{O}{iv} also contribute
to the radiative force. For the highest effective temperatures,
$T_\text{eff}\geq40\,000\,\text{K}$, the acceleration by the \ion{O}{v} and
\ion{O}{vi} lines for the scaled solar chemical composition is not compensated
for by acceleration by nitrogen for  the CNO-processed chemical composition,
which leads to a decrease of the mass-loss rate.

\paragraph{${Z=0.01\,Z_\odot}$}

For supergiant wind models at $Z=0.01\,Z_\odot$ and
$T_\text{eff}\leq35\,000\,\text{K}$ the change from the scaled solar chemical
composition to the CNO-processed chemical composition does not significantly
affect the line force, because instead of some oxygen lines there are new strong
nitrogen lines that are able to significantly accelerate the wind. For the
higher effective temperature $\boldsymbol T_\text{eff}=37\,500\,\text{K}$ and
the CNO-processed chemical composition, instead of the \ion{O}{v} lines for the
scaled solar chemical composition, the \ion{N}{v} resonance doublet becomes
optically thick. This leads to an increase of the radiative force, because the
\ion{N}{v} resonance doublet is located in the spectral region with higher flux.
A similar effect also occurs in model 350-3.

\begin{figure}
\centering
\resizebox{\hsize}{!}{\includegraphics{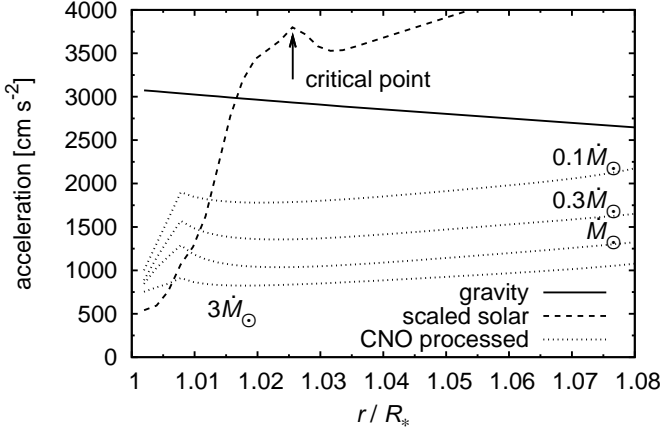}}
\caption{Radial variations of the absolute values of gravitational and
radiative acceleration in model 300-3
and $Z=0.01\,Z_\odot$ (see Table \ref{venzettab}).
The radiative acceleration  for a scaled solar chemical composition is
plotted as the dashed line.
Since there is no wind for the CNO-processed scaled solar chemical
composition, the radiative acceleration for several test models is overplotted.
The test models were calculated for four different values of the
mass-loss rate denoted in the graph.
($\dot M_\odot$ is the mass-loss rate for the scaled solar chemical
composition -- see text). The critical point of the 300-3 model
with $Z=0.01\,Z_\odot$ is also denoted in the graph.}
\label{3003sil}
\end{figure}

\paragraph{Cases without wind}

We were not able to converge our models for some stars with a CNO-processed
chemical composition. For $Z=0.03\,Z_\odot$ this occurred only for model 300-5,
but for $Z=0.01\,Z_\odot$ this occurred for models 300-3, 300-5, 350-5, 375-3,
375-5, and all models with $T_\text{eff}\geq40\,000\,\text{K}$. The failed
convergence indicates that there is not enough radiative force to drive a wind
in this case. To verify this, we calculated additional models with a fixed
hydrodynamical structure, where we compared the radiative force with the gravity
\citep [as in] []{cnovit}. These models were calculated for four different fixed
mass-loss rates equal to $3\dot M_\odot$, $\dot M_\odot$, $\dot M_\odot/3$, and
$0.1\dot M_\odot$, where $\dot M_\odot$ is the mass-loss rate for a scaled solar
chemical composition and the same stellar parameters. In all these models the
radiative force was significantly lower than the gravity force (see
Fig.~\ref{3003sil} for an example of such calculations for a model 300-3). This
supports the conclusion that the total radiative force is not able to drive a
wind and that there is no wind for a CNO-processed chemical composition in the
mentioned cases.

\section{Metallicity variations and failure of the wind-momentum luminosity
relationship for low metallicities}

\begin{figure}
\centering
\resizebox{\hsize}{!}{\includegraphics{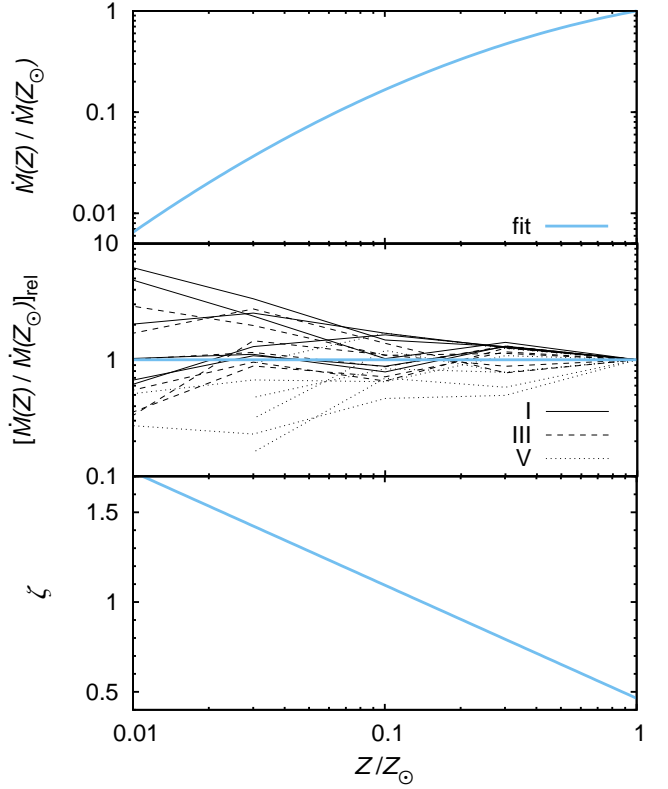}}
\caption{Variation of mass-loss rate with metallicity for the scaled solar chemical
composition.
{\em Upper panel}: 
mean dependence of the ratio of the scaled solar chemical composition
mass-loss rate on the solar metallicity mass-loss rate
after Eq.~\eqref{zetrov}.
{\em Middle panel}:
ratio of the mass-loss rate of individual model stars to the solar
abundance mass-loss rate plotted relatively to the fit
from the upper panel (Eq.~\ref{zetrov}).
{\em Lower panel}: variations of the steepness of the
dependence of $\log \hzav{\dot M(Z)/\dot M(Z_\odot)}$ on $\log( Z/Z_\odot)$ (see
Eq.~\eqref{recanynadlabem}) with metallicity.
}
\label{venzet}
\end{figure}

\begin{figure}
\centering
\resizebox{\hsize}{!}{\includegraphics{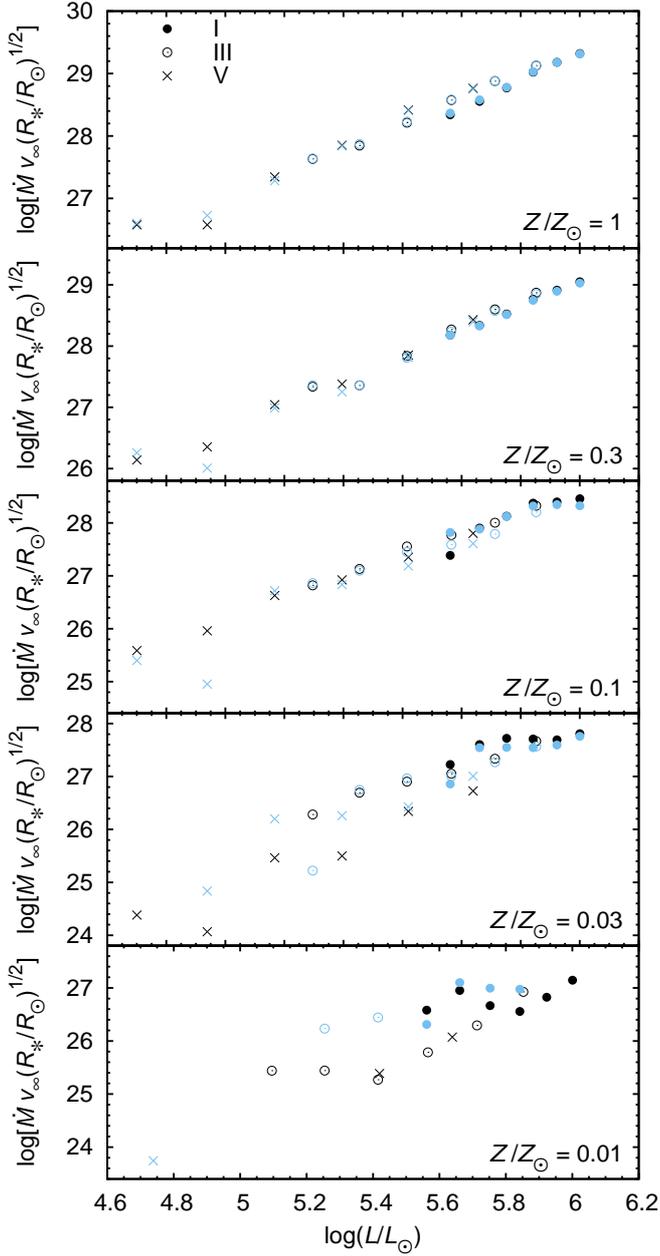}}
\caption{Modified wind momentum as a function of the stellar luminosity.
Different symbols correspond to different luminosity classes (full dots:
supergiants, empty dots: giants, crosses: main sequence stars). Black symbols
correspond to the scaled solar chemical composition and blue symbols to
the CNO-processsed scaled solar chemical composition.}
\label{momlu}
\end{figure}

The mass-loss rate generally decreases with decreasing metallicity, but the rate
of the decrease is not the same for all stars, as documented in
Fig.~\ref{venzet}. We fitted the dependence of the mass-loss rate for a given
star to the solar metallicity mass-loss rate $\dot M(Z)/\dot M(Z_\odot)$ by a
polynomial. An average of the logarithm of this ratio for all stars varies as
\begin{equation}
\label{zetrov}
\log \hzav{\dot M(Z)/\dot M(Z_\odot)}=0.46 \log( Z/Z_\odot)-0.31\hzav{\log(
Z/Z_\odot)}^2.
\end{equation}
For lower metallicities the decrease of the mass-loss rate with
metallicity is steeper than for higher metallicities \citep[see also, e.g.,]
[]{kudmet}. The metallicity variations are often expressed as $\dot M\sim Z^
\zeta$, where
\begin{equation}
\label{recanynadlabem}
\zeta
=\frac{\text{d}\log \hzav{\dot M(Z)/\dot M(Z_\odot)}}{\text{d} \log( Z/Z_\odot)}
=0.46-0.62\log( Z/Z_\odot)
.
\end{equation}
The dependence of $\zeta$ on the metallicity is given in the lower panel of
Fig.~\ref{venzet}. For solar metallicities, $ \zeta \approx0.5$ and we recover
the well-known dependence derived, for instance, by \citet{pusle}. For the Small
Magellanic Cloud metallicity we derive $ \zeta \approx0.7,$ in agreement with
\citet{nlteii}. By taking into account the range of metallicities considered by
\citet{vikolamet}, we can also explain the similar value $ \zeta =0.69$ derived
by \citet{vikolamet}. The ratio of the terminal velocity to the escape
velocity varies with metallicity on average as
$\varv_\infty/\varv_\text{esc}\sim (Z/Z_\odot)^{0.10}$, in agreement with the
results reported by \citet{kudmet}, for example. 

The middle panel of Fig.~\ref{venzet} demonstrates that the mass-loss rate
is not a simple function of the metallicity. Especially at
$Z/Z_\odot\lesssim0.1$ the metallicity dependence is sensitive to the stellar
parameters (as $\Teff$, $R_*$, and $M$). This is another argument in favor of
the conclusion \citep{lijana} that mass-loss rate predictions should be
calculated along with evolutionary calculations for actual stellar parameters.

For a given metallicity the modified wind momentum $\dot M \varv_\infty
\zav{R/{R}_{\odot}}^{1/2}$ depends mainly on the stellar luminosity \citep[and
references therein]{kupul}. With a proper metallicity calibration this
relationship may be used as a distance indicator. However, from Fig.~\ref{momlu}
it follows that the modified wind momentum uniquely depends on the stellar
luminosity only for stars with higher metallicities $Z/Z_\odot\gtrsim0.1$,
whereas for stars with lower metallicities this relationship shows a relatively
large scatter that makes any distance estimation problematic.

\section{Conclusions}

We studied the influence of the change of the chemical composition
caused by the CNO cycle on the hot star wind mass-loss rates.
This chemical composition evolution may result from the internal mixing
caused by the rotation.
The results were derived using our code, which enables calculation of a
stellar wind model with arbitrary chemical composition.

For metallicities $Z\gtrsim0.1 \, Z_\odot$ the change from the scaled solar
composition to the CNO-processed scaled solar composition does not influence the
mass-loss rate significantly. There are many strong CNO lines for these
metallicities, and the nitrogen lines are able to nearly completely compensate
for the missing line force coming from carbon or oxygen.

However, this is not true for the lowest abundances studied here, $Z\lesssim0.1
\, Z_\odot$. For these low metallicities the CNO elements significantly
contribute to the radiative force. As a result of the few optically thick lines,
the line force and therefore also the mass-loss rate is sensitive to the
individual fraction of CNO elements. Depending on the effective temperature, the
mass-loss rate may either decrease or increase when the chemical composition is
changed from the scaled solar one to the composition influenced by the CNO
cycle. 

The dependence of the mass-loss rate on the metallicity is significantly steeper
for $Z\lesssim0.1 \, Z_\odot$ than for the solar abundance. Moreover, the wind
mass-loss rate is very sensitive to the stellar parameters for low
metallicities. As a result, the wind momentum-luminosity relationship displays a
large scatter and is not unique, as it is at higher metallicities.

\begin{acknowledgements}
This research was supported GA\,\v{C}R  13-10589S.
Access to computing and storage facilities owned by parties and projects
contributing to the National Grid Infrastructure MetaCentrum, provided under the
program "Projects of Large Infrastructure for Research, Development, and
Innovations" (LM2010005), is greatly appreciated.
\end{acknowledgements}

\clearpage
\appendix
\onecolumn

\section{Adopted CNO abundances for the CNO-processed scaled solar chemical
composition and derived results for individual models}

\begin{table*}[h]
\caption{Abundances of CNO elements for CNO-processed scaled solar composition}
\centering
\begin{tabular}{cccccc}
\hline
\hline
$Z/Z_\odot$&1&0.3&0.1&0.03&0.01\\
\hline
$N_\text{C}/N_\text{H}$ & $2.41\times10^{-5}$ & $7.22\times10^{-6}$ & $2.41\times10^{-6}$ & $7.22\times10^{-7}$ & $2.41\times10^{-7}$\\
$N_\text{N}/N_\text{H}$ & $7.3 \times10^{-4}$ & $2.19\times10^{-4}$ & $7.3 \times10^{-5}$ & $2.19\times10^{-5}$ & $7.3 \times10^{-6}$\\
$N_\text{O}/N_\text{H}$ & $7.3 \times10^{-5}$ & $2.19\times10^{-5}$ & $7.3 \times10^{-6}$ & $2.19\times10^{-6}$ & $7.3 \times10^{-7}$\\
\hline
\end{tabular}
\label{cnoab}
\end{table*}

\begin{table*}[h]
\caption{Predicted wind parameters for individual stellar models. Mass-los rates
$\dot M$ are given in $M_\odot\,\text{year}^{-1}$ and terminal velocities
$\varv_\infty$ in $\text{km}\,\text{s}^{-1}$.}
\centering
\begin{tabular}{cccccccccccccccccccccccccccccc}
\hline
\hline
Model & $Z/Z_\odot$ & \multicolumn{2}{c}{Scaled solar} &
\multicolumn{2}{c}{CNO processed}\\
&&$\dot M$ & $\varv_\infty$&$\dot M$ & $\varv_\infty$\\
\hline
300-1 & 0.01 & $1.5\times10^{ -8}$ & 870 & $1.4\times10^{ -8}$ & 500\\
      & 0.03 & $4.1\times10^{ -8}$ & 1380 & $5.2\times10^{ -8}$ & 460\\
      & 0.10 & $8.0\times10^{ -8}$ & 1020 & $1.6\times10^{ -7}$ & 1390\\
      & 0.30 & $3.1\times10^{ -7}$ & 1620 & $3.3\times10^{ -7}$ & 1530\\
      & 1.00 & $4.7\times10^{ -7}$ & 1560 & $4.9\times10^{ -7}$ & 1580\\
      \hline
300-3 & 0.01 & $1.6\times10^{ -9}$ & 740 & \multicolumn{2}{c}{no wind}\\
      & 0.03 & $6.3\times10^{ -9}$ & 1330 & $7.1\times10^{-10}$ & 1030\\
      & 0.10 & $1.6\times10^{ -8}$ & 1820 & $2.0\times10^{ -8}$ & 1620\\
      & 0.30 & $4.6\times10^{ -8}$ & 2050 & $5.2\times10^{ -8}$ & 1960\\
      & 1.00 & $8.7\times10^{ -8}$ & 2150 & $9.3\times10^{ -8}$ & 2070\\
      \hline
300-5 & 0.01 & \multicolumn{2}{c}{no wind} & \multicolumn{2}{c}{no wind}\\
      & 0.03 & $1.6\times10^{-10}$ & 920 & \multicolumn{2}{c}{no wind}\\
      & 0.10 & $1.3\times10^{ -9}$ & 1840 & $1.2\times10^{ -9}$ & 1310\\
      & 0.30 & $3.4\times10^{ -9}$ & 2510 & $5.2\times10^{ -9}$ & 2160\\
      & 1.00 & $9.3\times10^{ -9}$ & 2510 & $8.7\times10^{ -9}$ & 2860\\
      \hline
325-1 & 0.01 & $3.4\times10^{ -8}$ & 910 & $3.2\times10^{ -8}$ & 1350\\
      & 0.03 & $1.0\times10^{ -7}$ & 1340 & $1.2\times10^{ -7}$ & 1040\\
      & 0.10 & $2.1\times10^{ -7}$ & 1320 & $1.9\times10^{ -7}$ & 1350\\
      & 0.30 & $5.1\times10^{ -7}$ & 1460 & $5.1\times10^{ -7}$ & 1450\\
      & 1.00 & $8.4\times10^{ -7}$ & 1460 & $8.1\times10^{ -7}$ & 1610\\
      \hline
325-3 & 0.01 & $2.0\times10^{ -9}$ & 610 & $7.4\times10^{ -9}$ & 1000\\
      & 0.03 & $1.8\times10^{ -8}$ & 1210 & $1.7\times10^{ -8}$ & 1400\\
      & 0.10 & $4.1\times10^{ -8}$ & 1420 & $4.4\times10^{ -8}$ & 1240\\
      & 0.30 & $6.5\times10^{ -8}$ & 1530 & $7.2\times10^{ -8}$ & 1380\\
      & 1.00 & $1.8\times10^{ -7}$ & 1710 & $1.9\times10^{ -7}$ & 1730\\
      \hline
325-5 & 0.01 & \multicolumn{2}{c}{no wind} & $7.5\times10^{-11}$ & 430\\
      & 0.03 & $1.7\times10^{-10}$ & 400 & $3.5\times10^{-10}$ & 1150\\
      & 0.10 & $2.9\times10^{ -9}$ & 1800 & $2.6\times10^{ -9}$ & 200\\
      & 0.30 & $5.4\times10^{ -9}$ & 2440 & $7.1\times10^{ -9}$ & 830\\
      & 1.00 & $1.5\times10^{ -8}$ & 1490 & $1.9\times10^{ -8}$ & 1610\\
      \hline
350-1 & 0.01 & $1.8\times10^{ -8}$ & 910 & $2.3\times10^{ -8}$ & 1530\\
      & 0.03 & $1.3\times10^{ -7}$ & 1470 & $6.7\times10^{ -8}$ & 1850\\
      & 0.10 & $3.8\times10^{ -7}$ & 1220 & $3.4\times10^{ -7}$ & 1340\\
      & 0.30 & $8.1\times10^{ -7}$ & 1440 & $8.0\times10^{ -7}$ & 1430\\
      & 1.00 & $1.4\times10^{ -6}$ & 1520 & $1.3\times10^{ -6}$ & 1630\\
      \hline
350-3 & 0.01 & $9.6\times10^{-10}$ & 820 & $8.3\times10^{ -9}$ & 1430\\
      & 0.03 & $2.4\times10^{ -8}$ & 1430 & $1.8\times10^{ -8}$ & 2200\\
      & 0.10 & $8.2\times10^{ -8}$ & 1860 & $7.3\times10^{ -8}$ & 1640\\
      & 0.30 & $1.8\times10^{ -7}$ & 1600 & $1.7\times10^{ -7}$ & 1610\\
      & 1.00 & $4.5\times10^{ -7}$ & 1560 & $4.4\times10^{ -7}$ & 1640\\
      \hline
350-5 & 0.01 & \multicolumn{2}{c}{no wind} & \multicolumn{2}{c}{no wind}\\
      & 0.03 & $1.6\times10^{ -9}$ & 1020 & $4.7\times10^{ -9}$ & 1860\\
      & 0.10 & $1.3\times10^{ -8}$ & 1800 & $1.1\times10^{ -8}$ & 2580\\
      & 0.30 & $2.7\times10^{ -8}$ & 2260 & $2.6\times10^{ -8}$ & 2090\\
      & 1.00 & $4.6\times10^{ -8}$ & 2650 & $4.4\times10^{ -8}$ & 2390\\

\hline
\end{tabular}\qquad
\begin{tabular}{cccccccccccccccccccccccccccccc}
\hline
\hline
Model & $Z/Z_\odot$ & \multicolumn{2}{c}{Scaled solar} &
\multicolumn{2}{c}{CNO processed}\\
&&$\dot M$ & $\varv_\infty$&$\dot M$ & $\varv_\infty$\\
\hline
375-1 & 0.01 & $9.2\times10^{ -9}$ & 1400 & $1.9\times10^{ -8}$ & 1810\\
      & 0.03 & $1.1\times10^{ -7}$ & 1670 & $4.9\times10^{ -8}$ & 2530\\
      & 0.10 & $6.2\times10^{ -7}$ & 1350 & $4.9\times10^{ -7}$ & 1520\\
      & 0.30 & $1.4\times10^{ -6}$ & 1490 & $1.3\times10^{ -6}$ & 1510\\
      & 1.00 & $2.3\times10^{ -6}$ & 1640 & $2.2\times10^{ -6}$ & 1720\\
      \hline
375-3 & 0.01 & $2.3\times10^{ -9}$ & 1110 & \multicolumn{2}{c}{no wind}\\
      & 0.03 & $3.2\times10^{ -8}$ & 1460 & $1.7\times10^{ -8}$ & 2560\\
      & 0.10 & $1.2\times10^{ -7}$ & 2070 & $7.7\times10^{ -8}$ & 2100\\
      & 0.30 & $5.5\times10^{ -7}$ & 1420 & $5.2\times10^{ -7}$ & 1400\\
      & 1.00 & $1.0\times10^{ -6}$ & 1570 & $9.7\times10^{ -7}$ & 1650\\
      \hline
375-5 & 0.01 & \multicolumn{2}{c}{no wind} & \multicolumn{2}{c}{no wind}\\
      & 0.03 & $1.1\times10^{ -9}$ & 1520 & $4.2\times10^{ -9}$ & 2270\\
      & 0.10 & $2.1\times10^{ -8}$ & 2040 & $1.0\times10^{ -8}$ & 3530\\
      & 0.30 & $5.0\times10^{ -8}$ & 2480 & $3.8\times10^{ -8}$ & 2410\\
      & 1.00 & $1.8\times10^{ -7}$ & 2000 & $1.8\times10^{ -7}$ & 1970\\
      \hline
400-1 & 0.01 & $1.3\times10^{ -8}$ & 1840 & \multicolumn{2}{c}{no wind}\\
      & 0.03 & $1.2\times10^{ -7}$ & 1490 & $5.6\times10^{ -8}$ & 2550\\
      & 0.10 & $4.4\times10^{ -7}$ & 2030 & $4.4\times10^{ -7}$ & 1820\\
      & 0.30 & $1.8\times10^{ -6}$ & 1610 & $1.8\times10^{ -6}$ & 1580\\
      & 1.00 & $3.0\times10^{ -6}$ & 1810 & $3.0\times10^{ -6}$ & 1820\\
      \hline
400-3 & 0.01 & $6.4\times10^{ -9}$ & 1250 & \multicolumn{2}{c}{no wind}\\
      & 0.03 & $6.3\times10^{ -8}$ & 1410 & $3.1\times10^{ -8}$ & 2470\\
      & 0.10 & $2.0\times10^{ -7}$ & 2090 & $1.4\times10^{ -7}$ & 1860\\
      & 0.30 & $1.1\times10^{ -6}$ & 1520 & $1.0\times10^{ -6}$ & 1520\\
      & 1.00 & $1.8\times10^{ -6}$ & 1720 & $1.8\times10^{ -6}$ & 1720\\
      \hline
400-5 & 0.01 & $1.1\times10^{ -9}$ & 1060 & \multicolumn{2}{c}{no wind}\\
      & 0.03 & $5.3\times10^{ -9}$ & 2000 & $4.8\times10^{ -9}$ & 2710\\
      & 0.10 & $4.9\times10^{ -8}$ & 2220 & $2.6\times10^{ -8}$ & 2910\\
      & 0.30 & $1.5\times10^{ -7}$ & 2360 & $1.5\times10^{ -7}$ & 2090\\
      & 1.00 & $6.3\times10^{ -7}$ & 2010 & $6.3\times10^{ -7}$ & 1950\\
      \hline
425-1 & 0.01 & $2.5\times10^{ -8}$ & 2100 & \multicolumn{2}{c}{no wind}\\
      & 0.03 & $1.5\times10^{ -7}$ & 1590 & $8.5\times10^{ -8}$ & 2490\\
      & 0.10 & $4.8\times10^{ -7}$ & 2190 & $3.9\times10^{ -7}$ & 1970\\
      & 0.30 & $2.3\times10^{ -6}$ & 1830 & $2.2\times10^{ -6}$ & 1810\\
      & 1.00 & $3.7\times10^{ -6}$ & 2100 & $3.7\times10^{ -6}$ & 2030\\
      \hline
425-3 & 0.01 & $1.9\times10^{ -8}$ & 1780 & \multicolumn{2}{c}{no wind}\\
      & 0.03 & $1.2\times10^{ -7}$ & 1530 & $6.9\times10^{ -8}$ & 2160\\
      & 0.10 & $4.2\times10^{ -7}$ & 2010 & $3.5\times10^{ -7}$ & 1830\\
      & 0.30 & $1.8\times10^{ -6}$ & 1670 & $1.7\times10^{ -6}$ & 1640\\
      & 1.00 & $2.9\times10^{ -6}$ & 1860 & $2.9\times10^{ -6}$ & 1830\\
      \hline
425-5 & 0.01 & $4.0\times10^{ -9}$ & 1330 & \multicolumn{2}{c}{no wind}\\
      & 0.03 & $3.0\times10^{ -8}$ & 810 & $1.4\times10^{ -8}$ & 3280\\
      & 0.10 & $1.3\times10^{ -7}$ & 2200 & $8.1\times10^{ -8}$ & 2270\\
      & 0.30 & $6.2\times10^{ -7}$ & 1980 & $5.7\times10^{ -7}$ & 1990\\
      & 1.00 & $1.2\times10^{ -6}$ & 2170 & $1.2\times10^{ -6}$ & 2100\\

\hline
\end{tabular}
\label{venzettab}
\end{table*}

\end{document}